# Electrostatic design of polar metal-organic framework thin films

*Giulia Nascimbeni,[1] Christof Wöll,[2] and Egbert Zojer[1,\*]*

[1]Institute of Solid State Physics, NAWI Graz, Graz University of Technology, Petersgasse 16, 8010 Graz, Austria

[2]Institute of Functional Interfaces (IFG), Karlsruhe Institute of Technology (KIT), Hermann-von-Helmholtz Platz-1, 76344 Eggenstein-Leopoldshafen, Germany

*e-mail: egbert.zojer@tugraz.at

Keywords: metal-organic frameworks, electronic structure, electrostatic design, density functional theory, work-function change, polar MOFs, bonding asymmetry

In recent years, optical and electronic properties of metal-organic frameworks (MOFs) have increasingly shifted into the focus of interest of the scientific community. Here, we discuss a strategy for conveniently tuning these properties through electrostatic design. More specifically, based on quantum-mechanical simulations, we suggest an approach for creating a gradient of the electrostatic potential within a MOF thin film exploiting collective electrostatic effects. With a suitable orientation of the polar apical linkers, the resulting non-centrosymmetric packing results in an energy staircase of the frontier electronic states reminiscent of the situation in a pin-photodiode. This 1-D gradient of the electrostatic potential causes a closure of the global energy gap and also shifts core-level energies by an energy equaling the size of the original band gap. The realization of such assemblies could be based on so-called pillared layer MOFs, grown in an oriented fashion on a solid substrate employing layer by layer growth techniques. In this context, the simulations provide guidelines regarding the design of the polar apical linker molecules that would allow the realization of MOF thin films with the (vast majority of the) molecular dipole moments pointing in the same direction.

# 1. Introduction

Metal-organic frameworks (MOFs) consist of metal-oxo nodes connected by di- or higher-topic organic linkers. They form crystalline, highly regular, and porous structures.[1] The high variability in possible node structures and linkers has resulted in the synthesis of tens of thousands of different MOF structures with hugely differing properties.[2] Relying on their highly porous structure, MOFs have traditionally been employed in areas like catalysis [3–5] gas storage,[6–8] and gas separation.[9,10] More recently, also their optical and electronic properties have gained considerable interest,[11–15] resulting in applications like sensing [16,17] and light harvesting.[18–20] On more fundamental grounds, in recent years the dynamics of charge carriers[13,21–25] and excitons[26–29] in MOFs has attracted considerable interest.

For many of the envisioned (opto)electronic applications of MOFs strategies for designing their electronic properties would be of distinct relevance. A highly promising approach for locally manipulating the electronic structure is electrostatic design. It relies on the fabrication of structures containing periodic arrangements of polar entities.[30,31] The superposition of the fields of the periodically arranged dipoles result in so-called collective (also termed cooperative) electrostatic effects, which are commonly observed at organic-inorganic hybrid interfaces.[32–36] They originate from the fact that extended 2D layer of dipoles rigidly shift the electrostatic energy of electrons between the regions above and below the layers with the magnitude of the effect being proportional to the dipole density.[32–36]

Consequently, arranging dipoles in multiple, consecutive layers into an asymmetric structure, as depicted in **Figure 1**a, results in an energy staircase. This is schematically shown in Figures 1b and 1c for a model thin film containing four layers of point dipoles. Here, Figures 1b displays the position dependence of the electrostatic energy of an electron, $E_{elstat}$, as derived from the superposition of the electric fields of the four layers of point dipoles (including the divergence of the potential at the locations of the dipoles). Figure 1c highlights the expected impact of the dipole layers on the electronic states of a layer of semiconducting material sandwiched between the polar layers. More specifically, it describes the relative energetic shifts of the frontier electronic states, denoted as valence-band edge, VB (or highest occupied molecular orbital, HOMO) and conduction-band edge, CB (or lowest unoccupied molecular orbital, LUMO), respectively.

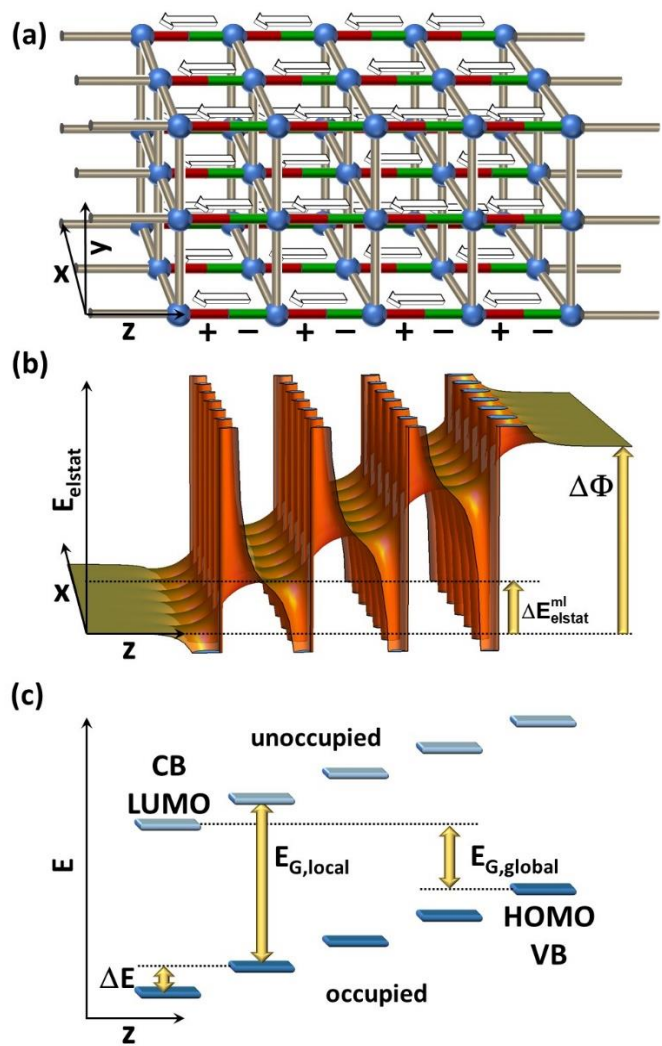

Figure 1: Design and properties of a polar metal-organic framework. Panel (a) shows the possible structure of a polar MOF in which the nodes (blue spheres) are connected by regular (apolar) linkers in x- and y-direction (grey cylinders) and by polar linkers in z-direction (red/green cylinders). The direction of the dipole moments is indicated by arrows. (b) Change in the electrostatic energy of an electron, $E_{elstat}$, induced by a series of four dipole sheets, which are infinitely extended in x- and y-direction. The dipoles in the sheets point in z-direction and they are arranged on a square lattice with periodicity, D. The distance between the dipole sheets also amounts to D and the plotting range to 6·D in x and z-

*directions. The actual value of D is inconsequential for the shape of the plot, as is the case also for the magnitude of the individual dipole moments. (c) Impact of the dipole layers on the frontier electronic states, which are assumed to be localized above, below, and between the dipole layers shown in panel (a). These could, for example, be the electronic states in the MOF localized on layers formed by nodes and apolar linkers, as described in the main text. There, also the various physical quantities contained in the plot are defined.*

In the current paper we employ fundamental electrostatic considerations and density-functional theory based band-structure calculations to predict the consequences of the inclusion of polar layers for the electronic properties of MOF thin films. Moreover, we will discuss a possible synthetic approach for realizing such systems.

## 2. Electronic structure of a polar MOF thin film

The fundamental impact of embedded polar layers on the properties of materials and interfaces can be derived from simple electrostatic considerations and has been described in detail, for example, in previous work.[36] In the following, we will briefly recapitulate the most relevant aspects and extend the considerations from [36] to systems containing multiple polar layers stacked on top of each other. We will refrain here from explicitly discussing the difference between stacked dipole layers and stacked individual dipoles and the artifacts arising from employing periodic boundary conditions, which has been discussed in an earlier publication.[37]

The stepwise change in electrostatic energy due to the presence of a single polar layer, $\Delta E_{elstat}^{ml}$, can be derived from the Poisson equation. It is given by:

$$\Delta E_{elstat}^{ml} = -\frac{q_e \mu}{\varepsilon_0 A} \approx \Delta E \qquad (1)$$

Here, $q_e$ is the charge of an electron, $\varepsilon_0$ is the vacuum permittivity, $\mu$ the component of each individual dipole moment orientated perpendicular to the layer, and $A$ is the area per dipole. Provided that the polar layers do not interfere with the intrinsic electronic properties of the semiconducting material between the layers, the frontier levels between successive semiconducting regions (e.g., layers formed by nodes and apolar linkers) are shifted by an energy, $\Delta E$, which amounts to $\Delta E = \Delta E_{elstat}^{ml}$ (see Figure 1c). In that situation, the local band gap within each region, $E_{G,local}$, remains unaffected by the polar layers.[37] Conversely, the global energy gap as the energetic difference between the highest unoccupied state and the lowest unoccupied state in the entire sample decreases.[37] From Figure 1c it can be inferred that this decrease amounts to:

$$E_{G,global} = E_{G,local} - n \cdot \Delta E \qquad (2)$$

In this context it has to be mentioned that in thermodynamic equilibrium the global gap has to remain ≥ 0 eV. Thus, when $n$ becomes ≥ $E_{G,local}/\Delta E$, electron transfer from the rightmost to the leftmost semiconducting region will occur in conjunction with a polarization of the material in between.[37,38] Alternative scenarios to establish thermodynamic equilibrium that have been observed for oxidic surfaces would be atomic rearrangements at the surface and adsorption/desorption processes that compensate for the dipole across the slab.[39,40] In the absence of thermodynamic equilibrium, the electrostatically triggered energy shifts can exceed the band gap. Indeed, for stacked polar molecules on surfaces energy shifts of up to 28 eV have been observed.[41]

From a practical point of view, one could envision to apply the energy staircase of the electronic levels in Figure 1c to guide the flow of charge carriers, to separate electrons and holes, or to dissociate excitons.[31] In that sense, the energetic staircase is somewhat reminiscent of the band diagram of a *pin*-junction typically used in photodetectors and solar cells.[42] There the linear position dependence of the band edges is not related to collective electrostatic effects. Rather, it originates from the field generated by uncompensated ionized dopants in the depletion regions of the p- and n-doped semiconductors that extends also into the intrinsic region of the junction. Still, the resulting driving force for separating charges is similar to the situation encountered here. The semiconducting elements necessary for (opto)electronic applications, which exploit the energy gradient could be directly built into MOF network.[12,13] Considering the highly porous nature of the MOF structures, one could, however, also think of first building the polar structure and then infiltrating it with (semi)conducting entities.[13,14,43]

In passing we note that the shift in electrostatic energy between successive semiconducting regions will not only affect the frontier levels, but will also shifts core-level binding energies, $BE_X^n$. When referencing them, e.g., to the Fermi-level of a (metallic) substrate (vide infra) one should again observe a shift with the number of polar layers, $n$, separating the probed atoms from the substrate. This yields the following expectation of the position dependence of $BE_X^n$:

$$BE_X^n = BE_X^0 - n \cdot \Delta E \qquad (3)$$

X here denotes the specific core level that is investigated (in the following discussion the $Zn_{2s}$ core level). Equation (3) suggest that, relying on the highly localized initial states, x-ray photoelectron spectroscopy can be used as an experimental tool for mapping the electrostatic shifts discussed here.[36,44,45] Finally, growing the above series of polar layers on a metallic substrate (see below) will change the work function of the substrate, with the net workfunction change, $\Delta \Phi$, amounting to (see Figure 1):

$$\Delta \Phi = n \cdot \Delta E_{elstat}^{ml} \qquad (4)$$

**3. Suggesting a strategy for realizing a polar MOF**

Conceptually, for realizing MOFs comprising a finite number of polar layers (as depicted in Figure 1a), three main criteria have to be fulfilled: (i) Polar linkers have to be incorporated in an oriented fashion, aligned along only one spatial direction, (ii) the (vast) majority of the linkers must be aligned with their dipoles pointing in the same direction yielding a non-centrosymmetric thin film, and (iii) one must be able to grow a finite, well defined number of layers comprising polar linkers, potentially sandwiched between layers in which the linkers are apolar.

Ideally suited for achieving especially criteria (i) and (iii) are layer-by-layer growth techniques[46–49] applied to surface-mounted MOFs (SURMOFs).[47,49] Here, the substrate is typically functionalized with suitable anchoring groups,[50] then exposed to a metal source (e.g., Zn-acetate), rinsed, exposed to one type of linkers, rinsed, potentially exposed to another type of linkers, rinsed etc. For growing polar MOFs, one needs nodes that bond differently in different directions; then, one has to introduce polar linkers in the step that triggers the growth of the MOF in the direction perpendicular to the substrate (the z-direction). Following this procedure, also heterolayers[51] can be produced, where in the present context varying between polar and apolar linkers along the z-direction (perpendicular to the substrate) would be particularly interesting.

For discussing the concept of electrostatically designing the energy landscape of MOFs we start from a prototypical layered-pillar SURMOF, built from so-called Zn-paddlewheels connected by 1,4-benzenedicarboxylates (BDC) (see **Figure 2** a,b). The resulting 2D planar structures are then connected, e.g., by pillars in the form of bipyridines. For realizing polar structures, the (symmetric) bipyridines need to be replaced by (asymmetric) polar analogues. One of the simplest possibilities is to replace the H atoms in 3 and 5 positions in one of the pyridines by fluorine atoms (m2F-BP, see Figure 2b).

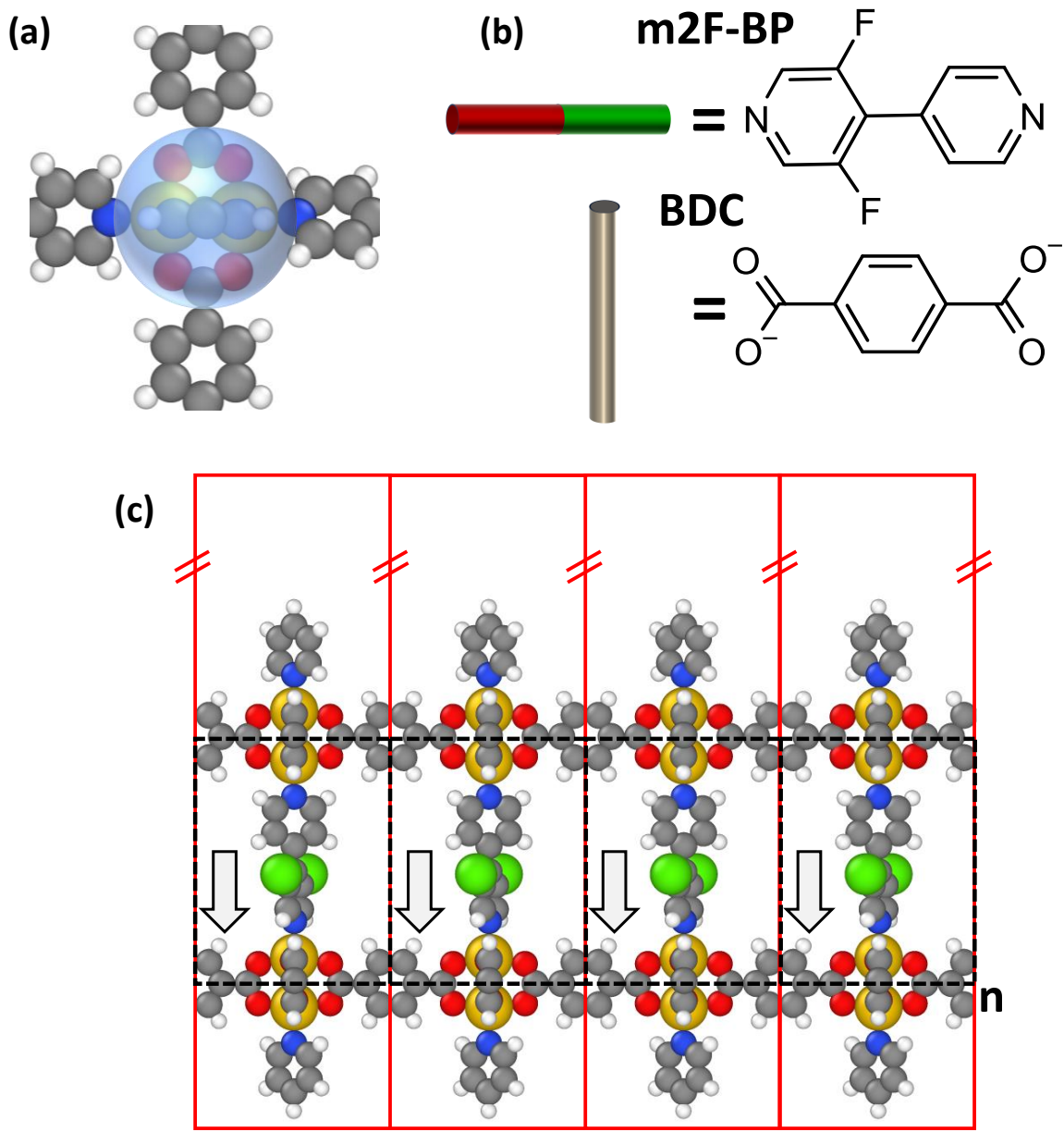

*Figure 2: (a) Chemical structure and bonding geometry of the Zn-paddlewheel node. The node consists of two Zn atoms connected by four carboxylic acid residues. Parts of the BDC linkers and the apical pyridine groups are also shown. The semitransparent blue sphere in the above plot corresponds to the blue sphere in the schematic structure shown in Figure 1a. (b) Chemical structures of the linkers: 1,4-benzenedicarboxylates (BDC) corresponds to the grey cylinders from Figure 1a and 3,5-difluoro-4,4'-bipyridine (with F in meta position, m2F-BP) corresponds to the red/green cylinders; (c) Structure of a polar monolayer consisting of one layer of polar m2F-BP molecules, two layers of BDC-linked Zn-paddlewheels, and pyridine layers at the top and bottom to saturate the paddlewheels. Each red*

*rectangle denotes the unit cell used in the band structure calculations (vide infra), the dashed black rectangles show the structure that is repeated when increasing the number of polar layers in the slab. The direction of the molecular dipoles is indicated by the arrows. Color code: grey: C; white: H; dark blue: N; red: O; yellow: Zn; green: F*

The major challenge in growing polar, non-centrosymmetric MOFs is to align all (or at least the vast majority) of the linkers such that their dipoles point in the same direction. From an electrostatic point of view, such a structure should be energetically unfavorable. Moreover, even if the dipole-dipole repulsion was weak due to the porous structure of the MOF (vide infra), entropy would favor a random orientation of the dipoles of the apical linkers. A possible strategy for overcoming these challenges would be to induce a large enough bonding asymmetry between the two pyridine docking sites in each linker molecule and the Zn-atoms, a topic that will be discussed in detail in section 4.2.

For simulating polar MOFs, it is necessary to model systems of finite thickness (as in a bulk calculation, the effect of the dipoles would vanish due to the 3D periodic boundary conditions).[37] Also from a practical point of view, such polar MOF thin films should be easier to realize than an actually polar bulk. Thus, as model systems we chose finite numbers of 2D periodic MOF layers consisting of BDC-linked Zn-paddlewheels, connected along the z-direction by m2F-BP linkers. These MOF thin films are then modelled employing the repeated slab approach, as described in the Methods section. As indicated in Figure 2c, such a slab contains n layers of polar m2F-BP molecules and n+1 layers of BDC-linked Zn-paddlewheels. The paddlewheels at the top and bottom of the slab are saturated by pyridine layers. Considering the lateral translational symmetry of the system, the unit cell can be chosen such that per layer it contains only one Zn-paddlewheel, two BDC linkers, and one m2F-BP molecule (see Supporting Information). In x- and y-direction, this unit cell is periodically repeated to represent the laterally quasi infinitely extended MOF.

## 4. Results and Discussion

### 4.1 Electronic structure of the polar MOF thin films

To demonstrate the key features of the electronic properties of polar MOF thin films, we first analyze the electrostatic energy landscape of a model system containing seven layers of properly aligned m2F-BP linkers. The corresponding plot of the electrostatic energy in the x-z plane half way between neighboring m2F-BP molecules is shown in **Figure 3**a for a thin film containing seven layers of polar linkers (n=7). The energy staircase is clearly visible, and is particularly well resolved when comparing the regions of the BDC-linked Zn-paddlewheels layers. A quantitative analysis reveals that the energy shift between adjacent layers amounts to ~0.26 eV per layer. An alternative way of visualizing the energy staircase would be to plot the dependence of the electrostatic energy averaged over planes perpendicular to the direction of the dipoles. The corresponding plots are shown in the Supporting Information, again for the n=7 situation and for a system consisting of a MOF thin film with four layers of polar apical linkers sandwiched between MOFs comprising two apolar apical linker layers on each side (i.e., the electrostatically generated "pin" junction mentioned above).

When comparing model systems with a varying number of polar layers, n, the energy steps result in a linear increase of the overall shift in electrostatic energy along the z-axis of the polar thin film, $\Delta\Phi$, with n (see Figure 3b). When such a thin film is grown on a metallic substrate with the z-direction normal to the surface, this results in an equivalent work-function shift as expressed by Eq. (4). Concomitantly, Figure 3c shows a linear decrease of the global gap with n, as predicted by Eq. (2). The observation of such a linear decrease instead of a 1/n dependence of the gap (like in conjugated oligomers and polymers)[52] confirms that the gap reduction is triggered by electrostatic effects rather than by an increasing conjugation. In passing we note that a similar observation has been made when calculating the properties of densely packed monolayers consisting of oligopyrimidine molecules.[37]

For simulations employing the PBE functional,[53,54] the above trend prevails up to 6 polar layers (n=6). For the 7$^{th}$ layer, the gap closes and, therefore, also the increase of $\Delta\Phi$ between n=6 and n=7 amounts to only 0.18 eV. As a consequence, the shift per layer is somewhat reduced from ~0.27 eV for n ≤ 6 to less than 0.26 eV for n=7. Note, that generalized gradient functionals like PBE substantially underestimate the band gap. As the result, the number of layers for which gap closure occurs will be different for different functionals. In order to improve the situation, we also performed calculations with the hybrid functional HSE06,[55] for which the underestimation of the gap is less serious. Unfortunately, with HSE06 in conjunction with periodic boundary conditions one encounters sharply increased computational costs. As a result, we only considered MOF thin films containing up to three polar layers. Nevertheless, extrapolating the HSE data from Figure 3c suggests that gap closure for the more realistic hybrid functionals will take place at around n=12. In passing we note that the somewhat steeper increase (decrease) of $\Delta\Phi$ ($E_{G,global}$) in the HSE06 calculations compared to the PBE results is the consequence of a larger dipole moment of the m2F-BP molecule for the former functional.

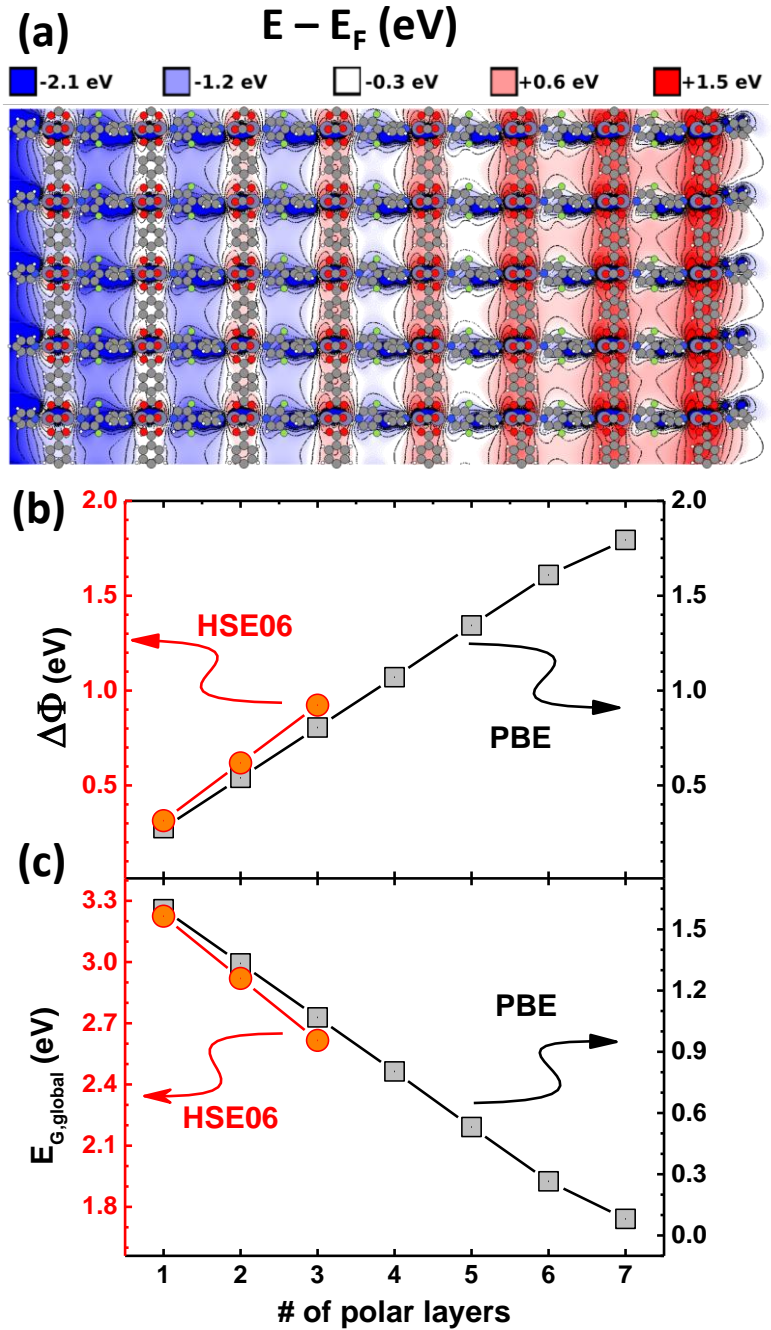

Figure 3: (a) evolution of the DFT-calculated electrostatic energy of an electron for a MOF thin film containing seven layers of polar, apical m2F-BP linkers. The graph shows the energy in a plane parallel to the x- and z-axes with the y-position chosen half-way between lines of linker molecules. (b) and (c) show the evolution of the overall step in electrostatic energy (the work-function change) and the global energy gap as a function of the number of layers of polar, apical m2F-BP linkers. The PBE results are plotted in black (right axes) and the HSE06 results in red (left axes).

In order to analyze local gaps and the energetic shifts between consecutive layers, we calculated the densities of electronic states projected onto the respective sub-systems.[37] These are shown for the valence region of the PBE calculations for the n=7 system in **Figure 4**a (with a plot covering a wider energy range contained in the Supporting Information). Here projections onto layers of BDC-linked Zn-paddlewheels are denoted by uppercase letters and projections onto m2F-BP layers are denoted by lowercase letters. The calculated local densities of states of adjacent equivalent layers have the same overall shape and are rigidly shifted by ~0.26 eV relative to each other. As discussed in section 2, this can be attributed to the shifts in electrostatic energies (see Figure 1c). When comparing the chemically different BDC-linked Zn-paddlewheels layers and the layers consisting of m2F-BP linkers (e.g., layers A and a in Figure 4) one observes a difference in the onsets of the occupied (unoccupied) DOSs, which amounts to 1.75 eV (1.19 eV). Overall, despite the decreasing global gap discussed above, the local gaps in the different sub-systems remains constant throughout the entire thin films. It amounts to 3.04 eV in the BDC-linked Zn-paddlewheels layers and to 3.64 eV in the m2F-BP layers. The HOMO (VB-edge) is localized on the topmost BDC-linked Zn-paddlewheels layer (layer H), while at least in the PBE calculations the LUMO (CB-edge) is calculated to be in the lowest m2F-BP layer (layer a).

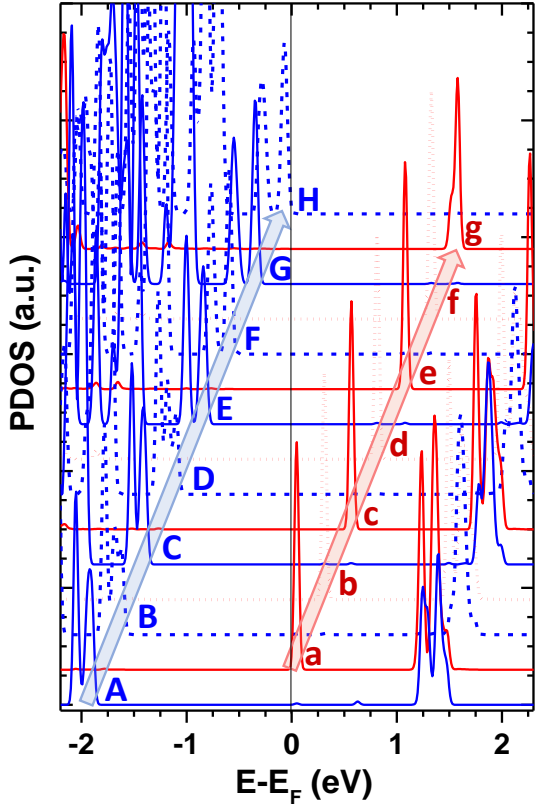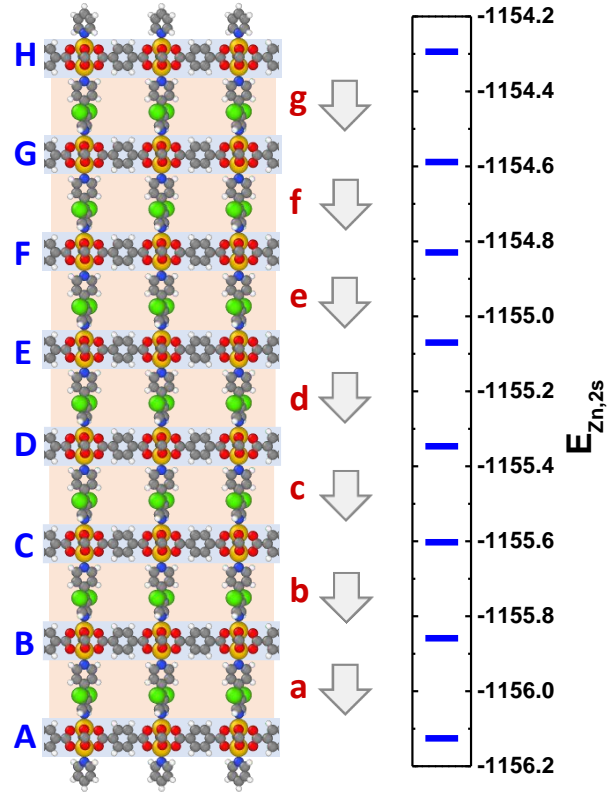

Figure 4: (a) DFT-calculated density of states projected onto the m2F-BP layers (red curves, denoted by small letters) and onto BDC-linked Zn-paddlewheels layers (blue curves, denoted by capital letters) for the n=7 thin film. The naming of the layers is shown in the central panel. (b) $Zn_{2s}$ core level energies for successive BDC-linked Zn-paddlewheels layers (the plot contains the average value for the two Zn-atoms in one node; for additional details see main text).

As a last aspect, we assess the impact of the polar layers on the core-level binding energies, as (via XPS) these should serve as an experimentally accessible probe for the shifts in electrostatic energy.[56] To that aim, Figure 4b show the binding energies of the $Zn_{2s}$ core levels for the 7-layer system calculated within the initial state approach.[57–60] Core-level shifts are typically well reproduced by this approach,[61–65] even though the absolute core-level energies need to be rigidly sifted to be compared to experiments (as in the simulations they are associated with the energies of Kohn-Sham orbitals, screening is neglected). The calculated overall shift between the Zn-atoms in the top and bottom layers amounts to as much as 1.83 eV, which (again) corresponds to ~0.26 eV per layer. This implies that the effect described here should be easily identified in XP spectra. Moreover, provided that for a specific system the number of polar layers was known (e.g., from the SURMOF production process), the actually observed XPS shift would yield information on the degree of alignment of the dipoles of the apical linkers.

The above results show that the expectations from the purely electrostatic considerations in section 2 are fully met by the results of the quantum-mechanical calculations. One of the reasons why the simple models from section 2 work out so well here, is that the BDC and m2F-BP linkers are quantum-mechanically decoupled from each other by the nodes, i.e., the different subsystems do not interact strongly with. Thus, the electronic states can be efficiently shifted relative to each other by the variations in the electrostatic potential. After discussing the expected electronic structure of polar MOF thin films, we next address, how simulations can help addressing the challenges one will encounter in assembling ordered, all-parallel, non-centrosymmetric arrangements of the polar linkers.

**4.2 Aligning the polar linkers during MOF growth – bonding asymmetry**

A common strategy for growing polar layers at interfaces is the use of anchoring groups, like thiolates on Au. In such systems, each molecule typically contains a single thiol group.[66–69] Its bonding to the substrate aligns the molecules provided that the bonding strength is large enough to overcome the dipole-dipole repulsion effects between adjacent moieties. Compared to polar, tightly packed, thiolate-bonded monolayers on surfaces, the dipole density is strongly reduced in highly porous MOFs. As a result, dipole-dipole repulsion effects are substantially reduced (vide infra) and, thus, should loosen the requirements for the docking groups. Still, they need to bond strongly enough to stabilize the MOF structure and weakly enough to break and reform in the self-assembly processes during MOF growth. The main challenge is, however, that in order to grow 3D structures, the linkers need to contain two docking functionalities. A possible strategy for still aligning the dipoles could be layer-by-layer growth of the MOF (vide supra) in combination with asymmetric linkers containing two distinctly different docking groups, one of which binds to the nodes with significantly higher bonding strength than the other.

To put this discussion on a more quantitative level, we first calculated bonding energies and their asymmetries for several chemically related linkers with regard to the pillared layer MOF-types discussed above. To separate the impact of the bonding from electrostatic interactions between neighboring polar linkers, the bonding energies were first calculated for monomer systems consisting of only a single (saturated) Zn-paddlewheel unit as shown in **Figure 5**a. In the following discussion, "up" and "down" refer to the orientation of the linker dipole. The systems considered in addition to m2F-BP are shown in Figure 5b and all simulated monomer structures can be found in the Supporting Information. **Table 1** contains the dipole moments of the isolated molecules and of the "up" and "down" monomer. Also the bonding energies and their asymmetries in the above-mentioned monomer systems are listed. Bonding energies, $E_b$, are defined as the difference between the energies of the system with the linker bonded to the paddlewheel and of the two isolated subsystems (polar linker and paddlewheel monomer with the linker removed). For the determination of $E_b$, the geometries of all (sub)systems were optimized disregarding solvent effects.

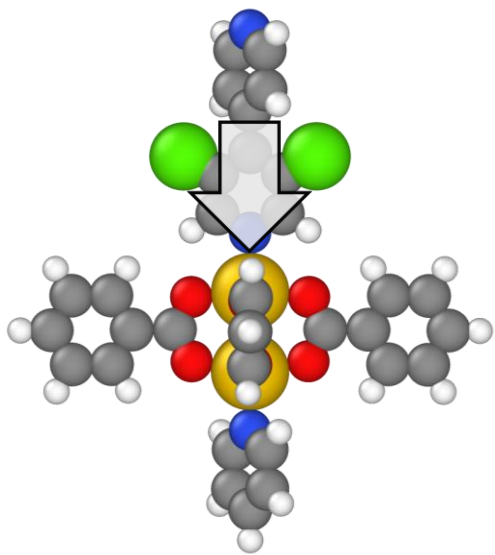

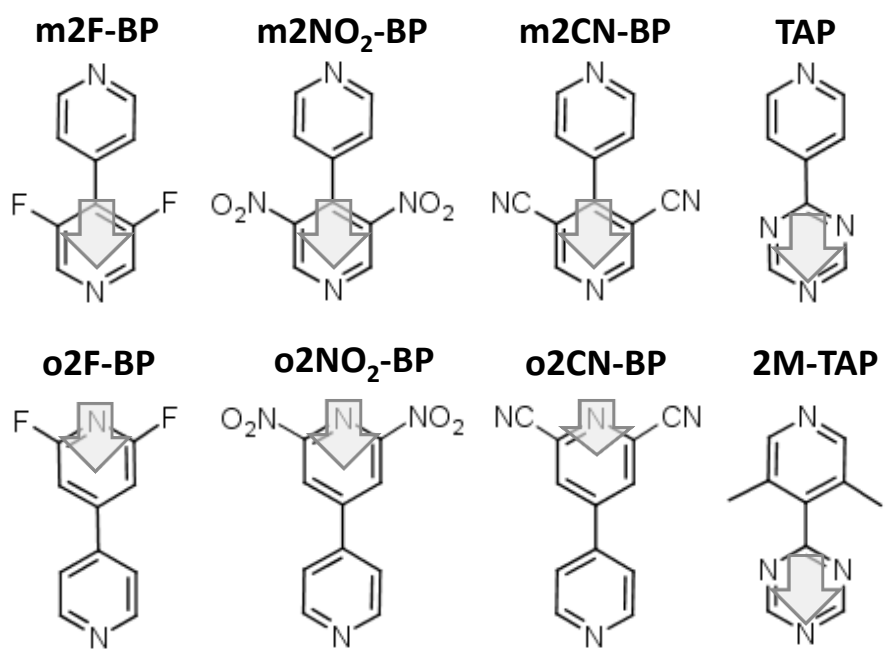

*Figure 5: (a) structure of a saturated Zn-paddlewheel monomer bonded to a polar apical linker (here m2F-BP) in the "down" orientation. (gray: C; whit: H; blue: N; red: O; yellow: Zn) (b) chemical structures of the polar apical linkers considered in the present study: 3,5-difluoro-4,4'-bipyridine (with F in meta position, m2F-BP), 2,6-difluoro-4,4'-bipyridine (with F in ortho position, o2F-BP), 3,5-dinitro-4,4'-bipyridine linker (m2NO$_2$-BP), 2,6-dinitro-4,4'-bipyridine linker (o2NO$_2$-BP), 3,5-dicyano-4,4'-bipyridine (m2CN-BP), 2,6-dicyano-4,4'-bipyridine (o2CN-BP), 4-s-triacinylpyridine (TAP), 3,5-dimetil-4-s-triacinylpyridine (inducing a twist, 2M-TAP).*

*Table 1: PBE-calculated values of the molecular dipole moments, $\mu_{mol}$ (they are obtained for a downward orientation of the dipoles and, thus, are reported as being negative); monomer-dipole moments for dipoles in the apical linkers in down orientation, $\mu_{down}$, (see Figure 5a) and up orientation, $\mu_{up}$; estimate for overall dipole of two saturated paddle-wheels connected by a polar linker, $\Delta\mu$ (see below); bonding energy between saturated paddle-wheel and apical linkers for dipole-down, $E_{b,down}$, and dipole-up, $E_{b,up}$, orientations; and bonding asymmetry, $\Delta E_b = E_{b,up} - E_{b,down}$; a negative value implies that the up-conformation is more strongly bonded. $\Delta\mu$ is given by $\Delta\mu = \mu_{up} - \mu_{down} - \mu_{mol}$. The expression can be rationalized by the system comprising the bond dipoles of linkers bonded in up- and down orientation (the latter with an inverted sign) plus the dipole of the polar molecule. Bond dipole here refers to the change in dipole between the bonded structure and an isolated molecule. The reason, why $\mu_{mol}$ has to be subtracted rather than added is that already both, $\mu_{up}$ and $\mu_{down}$ contain polar molecules, albeit in different orientations (for further details see Supporting Information).*

|  | m2F-BP | o2F-BP | m2NO$_2$-BP | o2NO$_2$-BP | m2CN-BP | o2CN-BP | TAP | 2M-TAP |
|---|---|---|---|---|---|---|---|---|
| $\mu_{mol}$ (D) | -0.76 | -1.86 | -1.92 | -4.85 | -2.24 | -5.09 | -1.33 | -1.95 |
| $\mu_{down}$ (D) | -3.57 | -4.27 | -5.95 | -7.75 | -6.20 | -8.05 | -4.35 | -5.02 |
| $\mu_{up}$ (D) | -1.40 | -0.63 | -0.49 |  | -0.32 | 0.67 | -0.67 | -0.16 |
| $\Delta\mu$ (D) | -1.41 | -1.78 | -3.54 |  | -3.64 | -3.63 | -2.36 | -2.91 |
| $E_{b,down}$ (meV) | 995 | 1057 | 943 | 1019 | 949 | 1028 | 946 | 955 |
| $E_{b,up}$ (meV) | 1061 | 590 | 1028 |  | 1022 | 674 | 1058 | 943 |
| $\Delta E_b$ (meV) | **-66** | **467** | **-85** |  | **-73** | **354** | **-112** | **12** |

The data for the isolated linker molecules in Table 1 show that m2F-BP has the smallest dipole moment of all considered apical linkers; the dipole moment of TAP is nearly twice as large and the dipole moments increase further for o2F-BP, 2M-TAP and m2CN-BP. Particularly large dipole moments are obtained for -CN and -NO$_2$ substituents in 2,6- (i.e., in ortho) positions. Considering the monomer dipoles in down orientation (i.e., the clusters containing also a saturated paddlewheel), one still observes the smallest value of $\mu_{down}$ for m2F-BP. The difference between molecular and cluster dipoles, however, varies considerably. This shows that the polarity of the Zn-N bond is significantly impacted by the substitution pattern of the apical linker. A consequence of this is, for example, that periodic monolayers of m2F-BP-linked MOFs have energetic steps per polar monolayer that are rather similar to their o2F-BP-linked counterparts (see Supporting Information), despite the significantly different molecular dipoles of the linker molecules. This can be rationalized by the estimates for the overall dipoles of two saturated paddle-wheels connected by a polar linker, $\Delta\mu$, in Table 1.

As far as the energetic stabilities of the N-Zn bonds are concerned, they are rather similar for all systems with the notable exceptions of $E_{b,up}$ for o2F-BP and o2CN-BP. The latter two are strongly reduced, which we attribute to steric repulsions between node and linker molecule due to the F atoms or -CN groups in ortho position. As a consequence, of all linkers shown in Figure 5, o2F-BP and o2CN-BP are the only ones with an appreciable asymmetry in the bonding energies, $\Delta E_b$. I.e. these two ortho-linked systems hold the highest promise for a spontaneous alignment of the polar linker. A possible complication when exploiting the above-mentioned steric effects is, however, that the reduced binding energies for the up-configuration might destabilize the entire structure. This becomes apparent for the o2NO$_2$-BP molecule attached to a Zn-paddlewheel in up orientation, where we calculated a strongly distorted structure with a broken N-Zn bond (see Supporting Information), which would prevent the formation of a 3D MOF.

A few additional aspects are worthwhile mentioning in the context of the energetic stability of aligned polar monolayers: (i) For the "up" version of the m2F-BP monomer we also calculated the total energy of the system as a function of the distance between the node and the m2F-BP linker. This yielded a monotonically increasing curve (see Supporting Information), indicating that there would be no activation energy barriers impacting the kinetics of bond formation and bond breaking.

(ii) As mentioned above, due to the reduced dipole density in the porous MOF, one can expect that dipole-dipole interactions within a linker layer should be strongly reduced compared to, e.g., a densely packed self-assembled monolayer on a surface. To quantify the effect, we recalculated the bonding energies for the m2F-BP system for a layer of BDC-linked Zn-paddlewheels saturated on one side with pyridines and on the other side with m2F-BP molecules comparing the situations for all m2F-BP molecules aligned in "up" and "down" orientations. Additionally, we constructed a layer with m2F-BP molecules aligned in an alternating, checkerboard-type fashion (see Supporting Information). From an electrostatic point of view, one would expect the checkerboard system to be clearly the most stable one, but the calculations yield the following order of average bonding energies per molecule: $E_{b,up}$=1108 meV > $E_{b,check}$=1082 meV > $E_{b,down}$=1048 meV. This means that for the monolayer the same bonding asymmetry is obtained as for the monomer calculations (60 meV vs. 66 meV in Table 1). The checkerboard arrangement of the m2F-BP molecules is essentially half way between the two other configurations ($E_{b,check}$ is only 4 meV higher than the average of $E_{b,up}$ and $E_{b,down}$). This implies that dipole-dipole repulsion plays a close to negligible role for the possible formation of a polar m2F-BP layer. To rationalize this a priori unexpected finding, one has to keep in mind that due to the porous nature of the MOF, the lateral distance between m2F-BP molecules is approximately 11 Å; moreover, the involved dipole moments are only moderately large. Indeed, a back of the envelope calculation reveals that the repulsion energy between two parallel dipoles of 0.76 Debye at a distance of 11 Å amounts to only 0.3 meV, which for a periodic arrangement of dipoles becomes 1.2 meV (employing a square Topping model [70]).

(iii) Likewise, the interaction energy between dipoles in consecutive m2F-BP layers is negligibly small (see Supporting Information). This is again a consequence of large inter-dipole distances in conjunction with the much more rapid drop of the electric field outside a periodic dipole layer compared to an isolated dipole.[32]

In this context it should, however, be stressed that even if dipole-dipole repulsion effects do not influence the formation of linker layers with aligned dipole moments, entropy will still favor a random orientation of the dipoles. I.e., one still needs to develop energetic driving forces (like the above-mentioned bonding asymmetry) for growing polar MOF thin films. In passing we note that an alternative approach could be the use of suitably chosen protecting groups, which ensure that, in a first step, only one of the bonding sites of the apical linker can connect to the Zn-paddlewheels.

## 5. Summary and Conclusions

Exploiting collective electrostatic effects, we propose a strategy for controlling the electronic structure of MOF thin films. In particular, we show, how the electronic states in a pillared layer SURMOF can be shifted between consecutive layers by using bipyridines as pillar linkers (connecting the BDC-linked Zn-paddlewheels) and by simultaneously substituting one of the pyridine units with electron-withdrawing F atoms. When achieving an alignment of all apical, asymmetric linkers with the dipole moments arranged in a parallel fashion, the observed energetic shift per layer amounts to ~0.27 eV. The resulting energy staircase of the frontier levels in successive semiconducting regions (e.g., BDC-linked Zn-paddlewheel layers) is reminiscent of the situation in a *pin*-photodiode, although the origin of the energy gradient is fundamentally different in both cases (field due to ordered 2D dipole layers vs. space charge regions due to uncompensated ionized dopants). The polar linker layers have virtually no impact on the local energy gaps of individual BDC-linked Zn-paddlewheel sheets. The global energy gap defined as the energetic difference of the highest occupied and lowest unoccupied states in the entire system, however, linearly with the number of polar layers. Eventually, for a sufficiently large number of polar layers, the global gap even vanishes. In addition to the positions of the frontier levels also the energies of the core levels are electrostatically shifted. Therefore, core-level spectroscopy appears as an ideal tool for probing the success of the alignment of the dipoles of the apical linkers.

A possible strategy for realizing that alignment would be to grow the MOFs in a layer-by-layer fashion, exploiting different binding energies of the two complexing units at the ends of the linkers, which are responsible for the docking to the MOF nodes. Exploring this approach by calculations for a variety of polar bipyridine derivatives reveals that the bonding asymmetries caused by polar substituents on only one of the rings become appreciable only when combining them with steric repulsion effects, as encountered for 2,6-substituted linkers. Such systems could, indeed, be promising, considering that dipole-dipole repulsion effects between aligned neighboring apical linkers are negligibly small due to the large inter-dipole distances. A possible complication in this context is, however, that the reduced binding energies for 2,6-bonded substituents might adversely affect the stability of the resulting MOFs suggesting that it would be worthwhile to also explore alternative approaches, like the use of protecting groups.

As an outlook, it should be noted that the lack of inversion symmetry in polar MOF thin films also makes them highly promising for first-order nonlinear optical (NLO) applications, with polar apical linkers acting as perfectly aligned NLO chromophores.[71] This should result in much larger nonlinearities compared to MOFs with centrosymmetric structures, which exploit second-order nonlinear effects.[72] In fact, the realization of polar MOFs could boost the applicability of MOFs in non-linear optics and might well help overcoming the challenge of properly aligning first-order NLO chromophores.[71] This would enable the use of MOFs in applications like second harmonic generation, or electrooptic switching.

**Methods:**

All calculations were performed using the FHI-aims code.[73–76] For many simulations periodic boundary conditions in conjunction with the repeated slab approach were applied. Periodic replicas of the slab were quantum-mechanically decoupled in z-direction by a 20 Å wide vacuum region. To decouple the slabs also electrostatically, a self-consistently determined dipole layer was included within the vacuum gap.[77,78] Following the results of convergence tests we chose 6×6×4 and 4×4×1 k-points grids for studying the bulk system (vide infra) and finite slabs, respectively. For the calculations we primarily employed the Perdew-Burke-Ernzerhof (PBE) functional,[53,54] which for determining binding energies was coupled to a Tkatchenko-Scheffler type van der Waals correction.[79] The latter had, however, no impact on the dipole moments, which determine the electronic structures of the systems, and also only weakly modified bonding asymmetries, as discussed in the Supporting Information. As far as the basis set is concerned, the FHI-aims default light settings, tier 2,[73] were used for every atom. A more detailed description of the used basis sets is again found in the Supporting Information. To assess the impact of the Basis Set Superposition Error when calculating bonding energies, we performed tests for the 3,5-difluoro-4,4'-bypiridine-based system increasing the basis-set size and employing a counterpoise correction. In both cases changes of bonding asymmetries were ≤ 1 meV, as shown in the Supporting Information. Convergence criteria for the self-consistency cycle were set to the default values for changes in the charge density ($10^{-5}$), the total energy ($10^{-6}$ eV), the forces ($10^{-4}$ eV·Å$^{-1}$). The geometry optimizations were performed using the version of the Broyden-Fletcher-Shanno-Goldfarb optimization algorithm enhanced by the trust radius method,[73] with a tolerance threshold of $10^{-2}$ eV·Å$^{-1}$. To determine the occupation of the Kohn-Sham eigenstate,s a Gaussian broadening function with a width of $\sigma=0.01$ eV was used. Only for the system with 7 layers of polar linkers the value had to be increased to 0.02 eV to reach convergence (due to the closing of the global gap). Test calculations on thinner layers with the increased broadening did not result in any appreciable changes of the electronic structure compared to the original 0.01 eV broadening. As the studied system contains Zn atoms, the atomic ZORA approximation was used[80] to account for relativistic effects. DOS and PDOS plots were obtained using the same Gaussian broadening function with a broadening of 0.01 eV as in the SCF procedure. The band gap (Kohn-Sham gap) as a function of n was obtained as the energy difference between the lowest unoccupied and the highest occupied states (for the above-described k-point sampling). Work function differences were derived from the energetic difference between the Fermi level and the vacuum level below and above the slab, respectively. Core-level energies were calculated within the initial-state approach[57–60] as energies of the $Zn_{2s}$ orbitals in the respective layers.

As generalized gradient functionals like PBE severely underestimate band gaps, to get somewhat improved values, the electronic structures of selected systems were recalculated using the hybrid Heyd-Scuseria-Ernzerhof (HSE06) exchange correlation functional.[55] These calculations were performed using a screening parameter of omega=0.11 Bohr$^{-1}$,[81] light settings and a tier 1 basis set with the further addition of the first two radial functions of the tier 2 basis set (a full tier 2 basis would have been computationally too expensive). The use of fewer basis functions for the HSE06 calculations is justified by convergence tests, which showed that there were no relevant differences between the results obtained with the full and the reduced tier 2 set. Due to the high computational costs of hybrid calculations combined with periodic boundary conditions, with HSE06 only single point calculations on the PBE optimized structures were performed. Plots of the electrostatic energy were produced using XCRYSDEN[82] and 3D geometries were plotted using OVITO.[83]

As far as the geometries of the investigated systems are concerned, an optimization of the bulk geometry of the di-zinc SBUs connected in the x,y-plane by terephthalic units and in the z direction by 3,5-difluoro-4,4'-bypiridine molecules yielded a tetragonal unit cell. This prompted us keep mutually orthogonal edges also for all derived structures. Starting from that bulk structure, as a first step the H atoms in position 3 and 5 in the bipyridine system were replaced by F atoms. The resulting structure was relaxed, simultaneously optimizing atomic positions and unit cell dimensions. Based on the optimized bulk structure, a monolayer slab was constructed, consisting of two layers of SBUs linked by BDC units and separated by a monolayer of 3,5-difluoro-4,4'-bypiridine molecules as apical linkers. The terminal Zn atoms were saturated with pyridines (see Figure 1c). Atomic positions and unit-cell dimensions in the x and y directions were optimized. As in this process the lengths of the unit-cell vectors did not change compared to the bulk, they were kept fixed for all further calculations. Systems with larger numbers of polar layers were constructed in an analogous fashion, where the atomic positions were fully optimized for systems containing up to 3 polar layers (n=3). As these optimizations yielded only very minor relaxations, thicker slabs were generated by replicating the m2F-BP layer and one of the central layers of BDC-linked Zn-paddlewheels without further geometry optimizations.

**Acknowledgements**

The authors are grateful to V. Obersteiner and T. Taucher for technical support in the calculations. Financial support by the Austrian Science Fund (FWF): P28051-N36 and by the Graz University of

Technology through a Lead Project (LP-03) is gratefully acknowledged. The quantum mechanical calculations have been performed using the Vienna Scientific Cluster (VSC3).

References

[1] H. Furukawa, K. E. Cordova, M. O'Keeffe, O. M. Yaghi, *Science* **2013**, *341*, DOI 10.1126/science.1230444.

[2] P. Z. Moghadam, A. Li, X.-W. Liu, R. Bueno-Perez, S.-D. Wang, S. B. Wiggin, P. A. Wood, D. Fairen-Jimenez, *Chemical Science* **2020**, *11*, DOI 10.1039/d0sc01297a.

[3] J. Liu, L. Chen, H. Cui, J. Zhang, L. Zhang, C. Y. Su, *Chemical Society Reviews* **2014**, *43*, 6011.

[4] L. Zhu, X. Q. Liu, H. L. Jiang, L. B. Sun, *Chemical Reviews* **2017**, *117*, 8129.

[5] V. Pascanu, G. González Miera, A. K. Inge, B. Martín-Matute, *Journal of the American Chemical Society* **2019**, *141*, 7223.

[6] M. Eddaoudi, J. Kim, N. Rosi, D. Vodak, J. Wachter, M. O'Keeffe, O. M. Yaghi, *Science* **2002**, *295*, 469.

[7] J. L. C. Rowsell, O. M. Yaghi, *Journal of the American Chemical Society* **2006**, *128*, 1304.

[8] L. J. Murray, M. Dinc, J. R. Long, *Chemical Society Reviews* **2009**, *38*, 1294.

[9] B. Chen, C. Liang, J. Yang, D. S. Contreras, Y. L. Clancy, E. B. Lobkovsky, O. M. Yaghi, S. Dai, *Angewandte Chemie - International Edition* **2006**, *45*, 1390.

[10] E. D. Bloch, W. L. Queen, R. Krishna, J. M. Zadrozny, C. M. Brown, J. R. Long, *Science* **2012**, *335*, 1606.

[11]  M. D. Allendorf, C. A. Bauer, R. K. Bhakta, R. J. T. Houk, *Chem. Soc. Rev.* **2009**, *38*, 1330.

[12]  I. Stassen, N. Burtch, A. Talin, P. Falcaro, M. Allendorf, R. Ameloot, *Chem. Soc. Rev.* **2017**, *46*, 3185.

[13]  L. Sun, M. G. Campbell, M. Dincă, *Angewandte Chemie International Edition* **2016**, *55*, 3566.

[14]  M. D. Allendorf, M. E. Foster, F. Léonard, V. Stavila, P. L. Feng, F. P. Doty, K. Leong, E. Y. Ma, S. R. Johnston, A. A. Talin, *J. Phys. Chem. Lett.* **2015**, *6*, 1182.

[15]  R. Haldar, L. Heinke, C. Wöll, *Adv. Mater.* **2020**, *32*, 1905227.

[16]  I. Stassen, N. Burtch, A. Talin, P. Falcaro, M. Allendorf, R. Ameloot, *Chemical Society Reviews* **2017**, *46*, 3185.

[17]  L. E. Kreno, K. Leong, O. K. Farha, M. Allendorf, R. P. Van Duyne, J. T. Hupp, *Chemical Reviews* **2012**, *112*, 1105.

[18]  H. J. Son, S. Jin, S. Patwardhan, S. J. Wezenberg, N. C. Jeong, M. So, C. E. Wilmer, A. A. Sarjeant, G. C. Schatz, R. Q. Snurr, O. K. Farha, G. P. Wiederrecht, J. T. Hupp, *Journal of the American Chemical Society* **2013**, *135*, 862.

[19]  T. Zhang, W. Lin, *Chemical Society Reviews* **2014**, *43*, 5982.

[20]  M. Oldenburg, A. Turshatov, D. Busko, S. Wollgarten, M. Adams, N. Baroni, A. Welle, E. Redel, C. Wöll, B. S. Richards, I. A. Howard, *Advanced Materials* **2016**, *28*, 8477.

[21]  S. Thomas, H. Li, R. R. Dasari, A. M. Evans, I. Castano, T. G. Allen, O. G. Reid, G. Rumbles, W. R. Dichtel, N. C. Gianneschi, S. R. Marder, V. Coropceanu, J.-L. Brédas, *Mater. Horiz.* **2019**, *6*, 1868.

[22]   A. J. Clough, N. M. Orchanian, J. M. Skelton, A. J. Neer, S. A. Howard, C. A. Downes, L. F. J. Piper, A. Walsh, B. C. Melot, S. C. Marinescu, *J. Am. Chem. Soc.* **2019**, *141*, 16323.

[23]   G. Skorupskii, B. A. Trump, T. W. Kasel, C. M. Brown, C. H. Hendon, M. Dincă, *Nature Chemistry* **2020**, *12*, 131.

[24]   M. Wang, M. Ballabio, M. Wang, H.-H. Lin, B. P. Biswal, X. Han, S. Paasch, E. Brunner, P. Liu, M. Chen, M. Bonn, T. Heine, S. Zhou, E. Cánovas, R. Dong, X. Feng, *J. Am. Chem. Soc.* **2019**, *141*, 16810.

[25]   L. S. Xie, G. Skorupskii, M. Dincă, *Chem. Rev.* **2020**, DOI 10.1021/acs.chemrev.9b00766.

[26]   Y. Tamai, H. Ohkita, H. Benten, S. Ito, *Journal of Physical Chemistry Letters* **2015**, *6*, 3417.

[27]   S. M. Menke, R. J. Holmes, *Energy and Environmental Science* **2014**, *7*, 499.

[28]   Q. Zhang, C. Zhang, L. Cao, Z. Wang, B. An, Z. Lin, R. Huang, Z. Zhang, C. Wang, W. Lin, *Journal of the American Chemical Society* **2016**, *138*, 5308.

[29]   M. Adams, M. Kozlowska, N. Baroni, M. Oldenburg, R. Ma, D. Busko, A. Turshatov, G. Emandi, M. O. Senge, R. Haldar, C. Wöll, G. U. Nienhaus, B. S. Richards, I. A. Howard, *ACS Applied Materials and Interfaces* **2019**, *11*, 15688.

[30]   B. Kretz, D. A. Egger, E. Zojer, *Advanced Science* **2015**, *2*, 1400016.

[31]   V. Obersteiner, A. Jeindl, J. Götz, A. Perveaux, O. T. Hofmann, E. Zojer, *Advanced Materials* **2017**, *29*, 1700888.

[32]   A. Natan, L. Kronik, H. Haick, R. T. Tung, *Advanced Materials* **2007**, *19*, 4103.

[33]   D. Cahen, R. Naaman, Z. Vager, *Advanced Functional Materials* **2005**, *15*, 1571.

[34]   G. Heimel, F. Rissner, E. Zojer, *Advanced Materials* **2010**, *22*, 2494.

[35]   O. L. A. Monti, *J. Phys. Chem. Lett.* **2012**, *3*, 2342.

[36]   E. Zojer, T. C. Taucher, O. T. Hofmann, *Adv. Mater. Interfaces* **2019**, *6*, 1900581.

[37]   F. Rissner, A. Natan, D. A. Egger, O. T. Hofmann, L. Kronik, E. Zojer, *Organic Electronics* **2012**, *13*, 3165.

[38]   J. Goniakowski, F. Finocchi, C. Noguera, *Rep. Prog. Phys.* **2007**, *71*, 016501.

[39]   O. Dulub, U. Diebold, G. Kresse, *Phys. Rev. Lett.* **2003**, *90*, 016102.

[40]   G. Kresse, O. Dulub, U. Diebold, *Phys. Rev. B* **2003**, *68*, 245409.

[41]   E. Ito, Y. Washizu, N. Hayashi, H. Ishii, N. Matsuie, K. Tsuboi, Y. Ouchi, Y. Harima, K. Yamashita, K. Seki, *Journal of Applied Physics* **2002**, *92*, 7306.

[42]   S. M. Sze, K. K. Ng, *Physics of Semiconductor Devices*, Wiley-Interscience, Hoboken, N.J, **2007**.

[43]   A. A. Talin, A. Centrone, A. C. Ford, M. E. Foster, V. Stavila, P. Haney, R. A. Kinney, V. Szalai, F. E. Gabaly, H. P. Yoon, F. Léonard, M. D. Allendorf, *Science* **2014**, *343*, 66.

[44]   T. C. Taucher, I. Hehn, O. T. Hofmann, M. Zharnikov, E. Zojer, *J. Phys. Chem. C* **2016**, *120*, 3428.

[45]   T. C. Taucher, E. Zojer, *Applied Sciences* **2020**, *10*, 5735.

[46]   O. Shekhah, H. Wang, S. Kowarik, F. Schreiber, M. Paulus, M. Tolan, C. Sternemann, F. Evers, D. Zacher, R. A. Fischer, C. Wöll, *J. Am. Chem. Soc.* **2007**, *129*, 15118.

[47]   D. Zacher, O. Shekhah, C. Wöll, R. A. Fischer, *Chem. Soc. Rev.* **2009**, *38*, 1418.

[48]  R. Makiura, S. Motoyama, Y. Umemura, H. Yamanaka, O. Sakata, H. Kitagawa, *Nature Materials* **2010**, *9*, 565.

[49]  J. Liu, C. Wöll, *Chem. Soc. Rev.* **2017**, *46*, 5730.

[50]  X.-J. Yu, Y.-M. Xian, C. Wang, H.-L. Mao, M. Kind, T. Abu-Husein, Z. Chen, S.-B. Zhu, B. Ren, A. Terfort, J.-L. Zhuang, *J. Am. Chem. Soc.* **2019**, *141*, 18984.

[51]  Z. Wang, J. Liu, B. Lukose, Z. Gu, P. G. Weidler, H. Gliemann, T. Heine, C. Wöll, *Nano Lett.* **2014**, *14*, 1526.

[52]  J. Gierschner, J. Cornil, H.-J. Egelhaaf, *Advanced Materials* **2007**, *19*, 173.

[53]  J. P. Perdew, K. Burke, M. Ernzerhof, *Phys. Rev. Lett.* **1996**, *77*, 3865.

[54]  J. P. Perdew, K. Burke, M. Ernzerhof, *Phys. Rev. Lett.* **1997**, *78*, 1396.

[55]  J. Heyd, G. E. Scuseria, M. Ernzerhof, *J. Chem. Phys.* **2003**, *118*, 8207.

[56]  I. Hehn, S. Schuster, T. Wächter, T. Abu-Husein, A. Terfort, M. Zharnikov, E. Zojer, *The Journal of Physical Chemistry Letters* **2016**, *7*, 2994.

[57]  P. S. Bagus, E. S. Ilton, C. J. Nelin, *Surface Science Reports* **2013**, *68*, 273.

[58]  J. P. Perdew, M. R. Norman, *Phys. Rev. B* **1982**, *26*, 5445.

[59]  R. Stowasser, R. Hoffmann, *J. Am. Chem. Soc.* **1999**, *121*, 3414.

[60]  D. P. Chong, O. V. Gritsenko, E. J. Baerends, *J. Chem. Phys.* **2002**, *116*, 1760.

[61]  M. Methfessel, V. Fiorentini, S. Oppo, *Phys. Rev. B* **2000**, *61*, 5229.

[62]  Y. Morikawa, T. Hayashi, C. C. Liew, H. Nozoye, *Surface Science* **2002**, *507–510*, 46.

[63]  G. Heimel, L. Romaner, J.-L. Brédas, E. Zojer, *Surface Science* **2006**, *600*, 4548.

[64]  N. P. Bellafont, F. Illas, P. S. Bagus, *Phys. Chem. Chem. Phys.* **2015**, *17*, 4015.

[65]   A. El-Sayed, P. Borghetti, E. Goiri, C. Rogero, L. Floreano, G. Lovat, D. J. Mowbray, J. L. Cabellos, Y. Wakayama, A. Rubio, J. E. Ortega, D. G. de Oteyza, *ACS Nano* **2013**, *7*, 6914.

[66]   A. Ulman, *Chem. Rev.* **1996**, *96*, 1533.

[67]   F. Schreiber, *Progress in Surface Science* **2000**, *65*, 151.

[68]   J. C. Love, L. A. Estroff, J. K. Kriebel, R. G. Nuzzo, G. M. Whitesides, *Chem. Rev.* **2005**, *105*, 1103.

[69]   C. Vericat, M. E. Vela, G. Benitez, P. Carro, R. C. Salvarezza, *Chem. Soc. Rev.* **2010**, *39*, 1805.

[70]   J. Topping, *Proceedings of the Royal Society A: Mathematical, Physical and Engineering Sciences* **1927**, *114*, 67.

[71]   S. R. Marder, B. Kippelen, A. K.-Y. Jen, N. Peyghambarian, *Nature* **1997**, *388*, 845.

[72]   C. Gu, H. Zhang, P. You, Q. Zhang, G. Luo, Q. Shen, Z. Wang, J. Hu, *Nano Lett.* **2019**, *19*, 9095.

[73]   V. Blum, R. Gehrke, F. Hanke, P. Havu, V. Havu, X. Ren, K. Reuter, M. Scheffler, *Computer Physics Communications* **2009**, *180*, 2175.

[74]   V. Havu, V. Blum, P. Havu, M. Scheffler, *Journal of Computational Physics* **2009**, *228*, 8367.

[75]   A. Marek, V. Blum, R. Johanni, V. Havu, B. Lang, T. Auckenthaler, A. Heinecke, H.-J. Bungartz, H. Lederer, *J. Phys.: Condens. Matter* **2014**, *26*, 213201.

[76]   V. W. Yu, F. Corsetti, A. García, W. P. Huhn, M. Jacquelin, W. Jia, B. Lange, L. Lin, J. Lu, W. Mi, A. Seifitokaldani, Á. Vázquez-Mayagoitia, C. Yang, H. Yang, V. Blum, *Computer Physics Communications* **2018**, *222*, 267.

[77]   J. Neugebauer, M. Scheffler, *Phys. Rev. B* **1992**, *46*, 16067.

[78]   C. Freysoldt, P. Eggert, P. Rinke, A. Schindlmayr, M. Scheffler, *Phys. Rev. B* **2008**, *77*, 235428.

[79]   A. Tkatchenko, M. Scheffler, *Physical Review Letters* **2009**, *102*, DOI 10.1103/PhysRevLett.102.073005.

[80]   E. van Lenthe, E. J. Baerends, J. G. Snijders, *J. Chem. Phys.* **1993**, *99*, 4597.

[81]   A. V. Krukau, O. A. Vydrov, A. F. Izmaylov, G. E. Scuseria, *J. Chem. Phys.* **2006**, *125*, 224106.

[82]   A. Kokalj, *Journal of Molecular Graphics and Modelling* **1999**, *17*, 176.

[83]   A. Stukowski, *Modelling and Simulation in Materials Science and Engineering* **2010**, *18*, 015012.

# Supporting Information

**Electrostatic design of polar metal-organic framework thin films**

Giulia Nascimbeni, Christof Wöll, and Egbert Zojer

## 1. Examples for studied unit cells

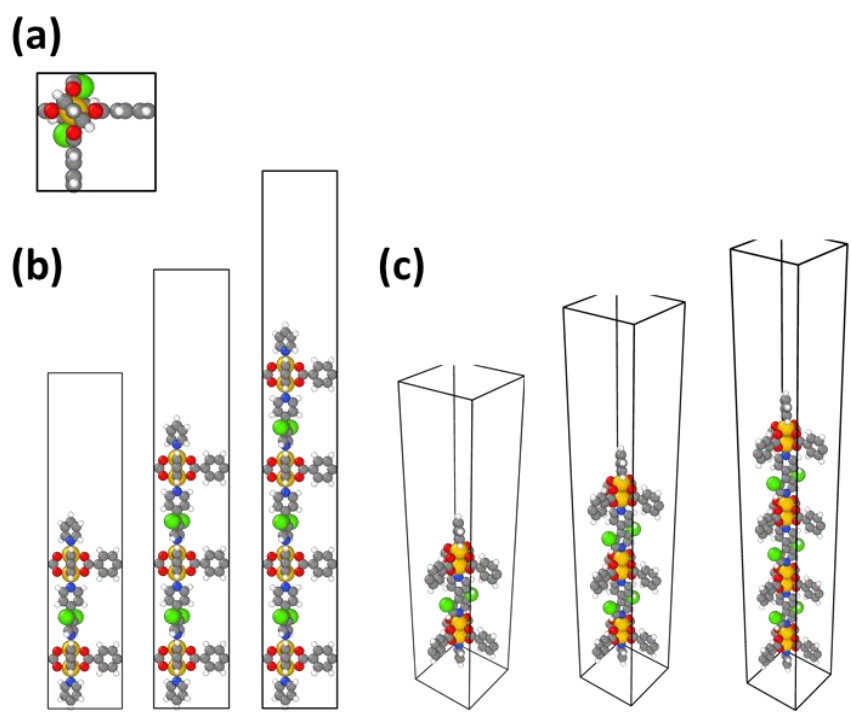

*Figure S1. Top (a), side (b), and diagonal (c) views of the unit cells used in the mono- bi- and trilayer Simulations (the top views are identical for all three systems) C: grey spheres; O: red spheres; H: small white spheres; F: green spheres, Zn: yellow spheres.*

## 2. Methodological aspects

Details on the used basis sets:

The basis functions employed in the FHI-aims simulations have the format

$$\Phi(r) = \frac{u(r)}{r} * Y_{lm}(\Theta, \Phi)$$

in spherical coordinates (r, Θ, Φ) relative to a given atomic center. FHI-aims provides for every atomic species a preconstructed *species_defaults* file. The used light basis sets were not further adjusted apart from adding tier 2 basis functions (see following table), because they afforded the required accuracy and efficiency.

Table S1. Basis functions that have been used for all calculations performed with FHI-aims[1]. The abbreviations read as follows[1]: X(nl, z), where X describes the type of basis function where H stands for hydrogen-like functions and ionic for a free-ion like radial function. The parameter n stands for the main/radial quantum number, l denotes the angular momentum quantum number (s, p, d, f, …), and z denotes an effective nuclear charge, which scales the radial function in the defining Coulomb potential for the hydrogen-like function. In the case of free-ion like radial functions, z specifies the onset radius of the confining potential. If auto is specified instead of a numerical value, the default onset is used.

|  | H | C | N | O | F | Zn |
|---|---|---|---|---|---|---|
| Minimal | valence (1s, 1.0) | valence (2s, 2.0) | valence (2s, 2.0) | valence (2s, 2.0) | valence (2s, 2.0) | valence (4s, 2.0) |
|  |  | Valence (2p, 2.0) | valence (2p, 3.0) | valence (2p, 4.0) | valence (2p, 5.0) | valence (3p, 6.0) |
|  |  |  |  |  |  | valence (3d, 10.0) |
| First tier | H(2s, 2.1) | H(2p, 1.7) | H(2p, 1.8) | H(2p, 1.8) | H(2p, 1.7) | H(2p, 1.7) |
|  | H(2p, 3.5) | H(3d, 6) | H(3d, 6.8) | H(3d, 7.6) | H(3d, 7.4) | H(3s, 2.9) |
|  |  | H(2s, 4.9) | H(3s, 5.8) | H(3s, 6.4) | H(3s, 6.8) | H(4p, 5.4) |
|  |  |  |  |  |  | H(4f, 7.8) |
|  |  |  |  |  |  | H(3d, 4.5) |
| Second tier | H(1s, 0.85) | H(4f, 9.8) | H(4f, 10.8) | H(4f, 11.6) | H(4f, 11.2) | H(5g, 10.8) |
|  | H(2p, 3.7) | H(3p, 5.2) | H(3p, 5.8) | H(3p, 6.2) | ionic (2p, auto) | H(H2p, 2.4) |
|  | H(2s, 1.2) | H(3s, 4.3) | H (1s, 0.8) | H (3d, 5.6) | H (1s, 0.75) | H(3s, 6.2) |
|  | H(3d, 7.0) | H(5g, 14.4) | H(5g, 16) | H(5g, 17.5) | H(4d, 8.8) | H(3d, 3) |
|  |  | H(3d, 6.2) | H(3d, 4.9) | H(1s, 0.75) | H(5g, 16.8) |  |

---

[1] As described in the FHI-aims manual, version January 23, 2017.

**Impact of the van der Waals correction:**

We tested the impact of the van der Waals corrections in the geometry optimizations and total energy calculations for the m2F-BP and o2F-BP-derived clusters and found that including them had no impact on the calculated dipoles (with changes for the up-, down- and molecular systems by < 0.02 Debye), as shown in Table S2. This strongly suggests that including van der Waals corrections has no impact on any of the reported electronic properties of the MOFs. Conversely, the binding energies, of course, increased significantly when including van der Waals corrections. The impact on the bonding asymmetry of up- vs. down configurations was, however, again rather minor. -66 (467) vs. -71 (488) meV for m2F-BP (o2F-BP) with and without van der Waals corrections (see Table S2).

Table S2: *Comparison of dipole moments ($\mu_{mol}$, $\mu_{down}$, $\mu_{up}$), binding energies ($E_{B,down}$, $E_{b,up}$) and binding energy asymmetries ($\Delta E_b$) of the m2F-BP and o2F-BP-derived clusters in their up- and down configurations and for the isolated molecules depending on whether or not a Tkatchenko-Scheffler type van der Waals correction[2] had been used in the geometry optimization process. Data plotted in red are also contained in the main manuscript. Data plotted in italics are for the o2F-BP apical linker.*

|  | m2F-BP vdW | *o2F-BP vdW* | m2F-BP no-vdW | *o2F-BP no-vdW* |
|---|---|---|---|---|
| $\mu_{mol}$ (D) | -0.76 | *-1.86* | -0.77 | *-1.86* |
| $\mu_{down}$ (D) | -3.57 | *-4.27* | -3.59 | *-4.26* |
| $\mu_{up}$ (D) | -1.40 | *-0.63* | -1.40 | *-0.64* |
| $E_{b,down}$ (meV) | 995 | *1057* | 763 | *830* |
| $E_{b,up}$ (meV) | 1061 | *590* | 833 | *341* |
| $\Delta E_b$ (meV) | **-66** | ***467*** | **-71** | ***488*** |

**Impact of basis-set superposition error:**

To test the role of the basis set for the calculation of binding energies (and especially for their asymmetries), we tested an extended (tier 3) basis set for the m2F-BP and o2F-BP systems and also performed counterpoise correction. As shown in Table S3, increasing the basis set size resulted in variations of the binding energies by a few meV (<< 1%). Most importantly, the change in the asymmetry of the binding energies for up and down configurations was only ≤ 1 meV. The same is observed for calculations employing a counterpoise correction (with somewhat larger changes for the absolute binding energies). Here, for technical reasons binding energies and molecular dipoles have been calculated for the geometries of all constituents fixed to the geometries they adopt in the cluster (i.e., geometry optimizations for constituents with the counterpoise correction switched on are not really sensible). Therefore, for assessing the impact of the counterpoise correction the two "tier2/fixed" columns for regular calculations at fixed geometries without counterpoise correction) need to be compared to the results of the calculations with counterpoise correction (last two columns of Table S3).

*Table S3: Comparison of dipole moments ($\mu_{mol}$, $\mu_{down}$, $\mu_{up}$), binding energies ($E_{B,down}$, $E_{b,up}$) and binding energy asymmetries ($\Delta E_b$) of the m2F-BP and o2F-BP-derived clusters in their up- and down configurations and for the isolated molecules as a function of the used basis sets. tier2/opt. are values obtained for simulations with the tier 2 basis set specified above in which not only the cluster geometries, but also the geometries of its isolated constituents (nodes and linkers) have been optimized; for tier3/opt. the same procedure has been applied but now including the full tier3 basis sets (again with light settings). tier2 fixed refers again to tier 2 calculations, but for calculating binding energies and molecular dipoles the linker molecule and node have been fixed to the geometries they adopt in the cluster. Finally, CP refers to values obtained with tier 2 calculations of linker and note (based on the cluster geometries), in which for the total energy calculations a counterpoise correction has been performed. Data plotted in red are also contained in the main manuscript. Data plotted in italics are for the o2F-BP apical linker.*

|  | m2F-BP tier2/opt. | o2F-BP tier2/opt. | m2F-BP tier3/opt. | o2F-BP tier3/opt. | m2F-BP tier2/fixed | o2F-BP tier2/fixed | m2F-BP CP | o2F-BP CP |
|---|---|---|---|---|---|---|---|---|
| $\mu_{mol}$ (D) | -0.76 | *-1.86* | -0.76 | -1.84 | -0.81 | -1.91 | - | - |
| $\mu_{down}$ (D) | -3.57 | *-4.27* | -3.57 | -4.24 | -3.57 | *-4.27* | - | - |
| $\mu_{up}$ (D) | -1.40 | *-0.63* | -1.43 | -0.64 | -1.40 | *-0.63* | - | - |
| $E_{b,down}$ (meV) | 995 | *1057* | 990 | *1052* | 1277 | *1367* | 1264 | *1353* |
| $E_{b,up}$ (meV) | 1061 | *590* | 1056 | *586* | 1389 | *885* | 1375 | *871* |
| $\Delta E_b$ (meV) | **-66** | ***467*** | **-66** | ***466*** | **111** | ***-482*** | **111** | ***-482*** |

## 3. Plane-averaged electrostatic energy

In addition to plotting the positional dependence of the electrostatic energy in an isodensity plot (as in Figure 3a of the main manuscript), one can also plot the electrostatic energy averaged over the xy-plane. This is shown for the 7-layer system (n=7) in Figure S2 and for the model of an analogue to a pin junction in Figure S3.

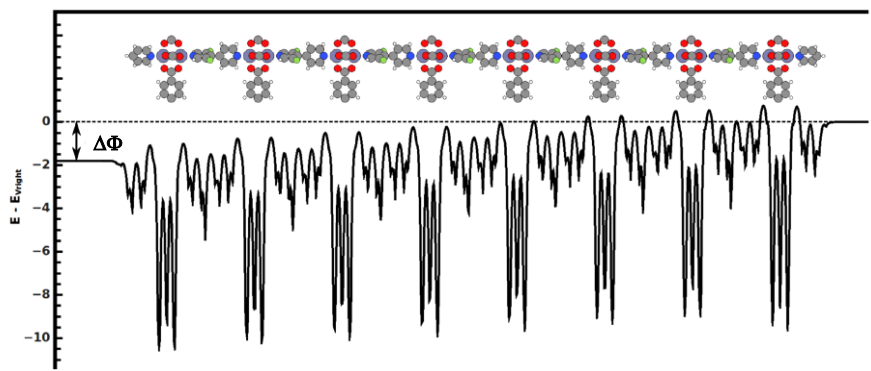

Figure S2. (a) PBE calculated, electrostatic energy averaged over the xy-plane and plotted as a function of the z-coordinate for the z-layer system. The energy is aligned to the right vacuum level. The energy difference $\Delta\Phi$ between the two sides of the sample is indicated by the arrow. The structure of the crystallographic basis is shown as an inset.

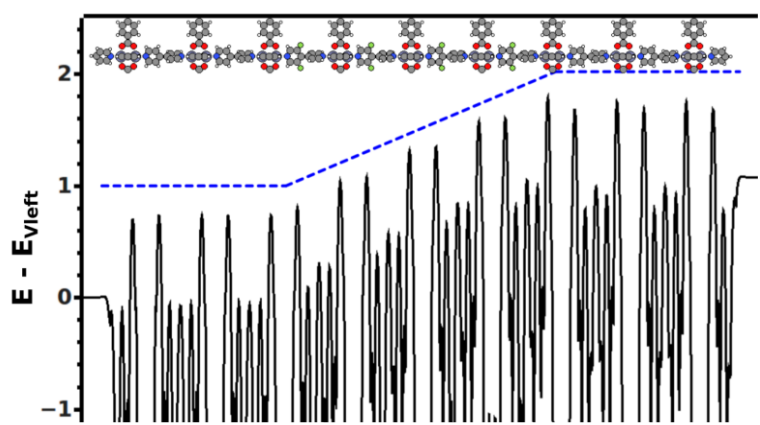

Figure S3. PBE calculated, electrostatic energy averaged over the xy-plane and plotted as a function of the z-coordinate for a system containing two layers of apolar 4,4'-bipyridine linkers, followed by four layers of polar 3,5-difluoro-4,4'-bipyridine linkers and finally by another two 4,4'-bipyridine linkers. The structure of the crystallographic basis is shown as an inset.

## 4. Density of states in the PBE calculations for the 7-layer system plotted over a wider energy range

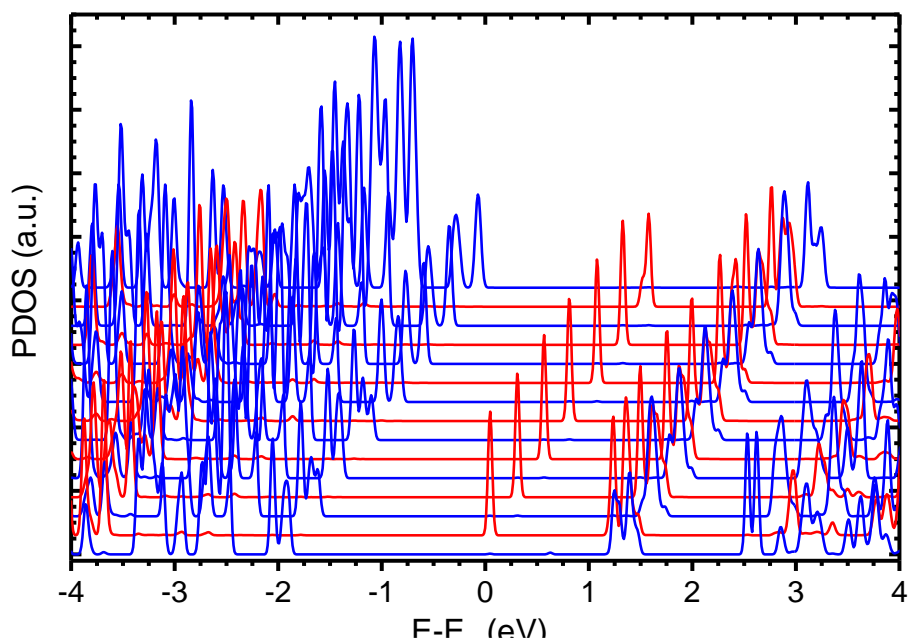

*Figure S4. Density of states for the 7-layer system projected onto successive BDC-linked Zn-paddlewheel layers (blue) and m2F-BP layers (red). In analogy to Figure 4a in the main manuscript, but plotted over a wider energy range.*

## 5. Structures of all "monomer" systems

To calculate the bonding energies of the polar apical linkers and their asymmetries, we simulated various linkers bonded to a single, saturated Zn-paddlewheel. The optimized structures of all considered systems are shown in Figure S6.

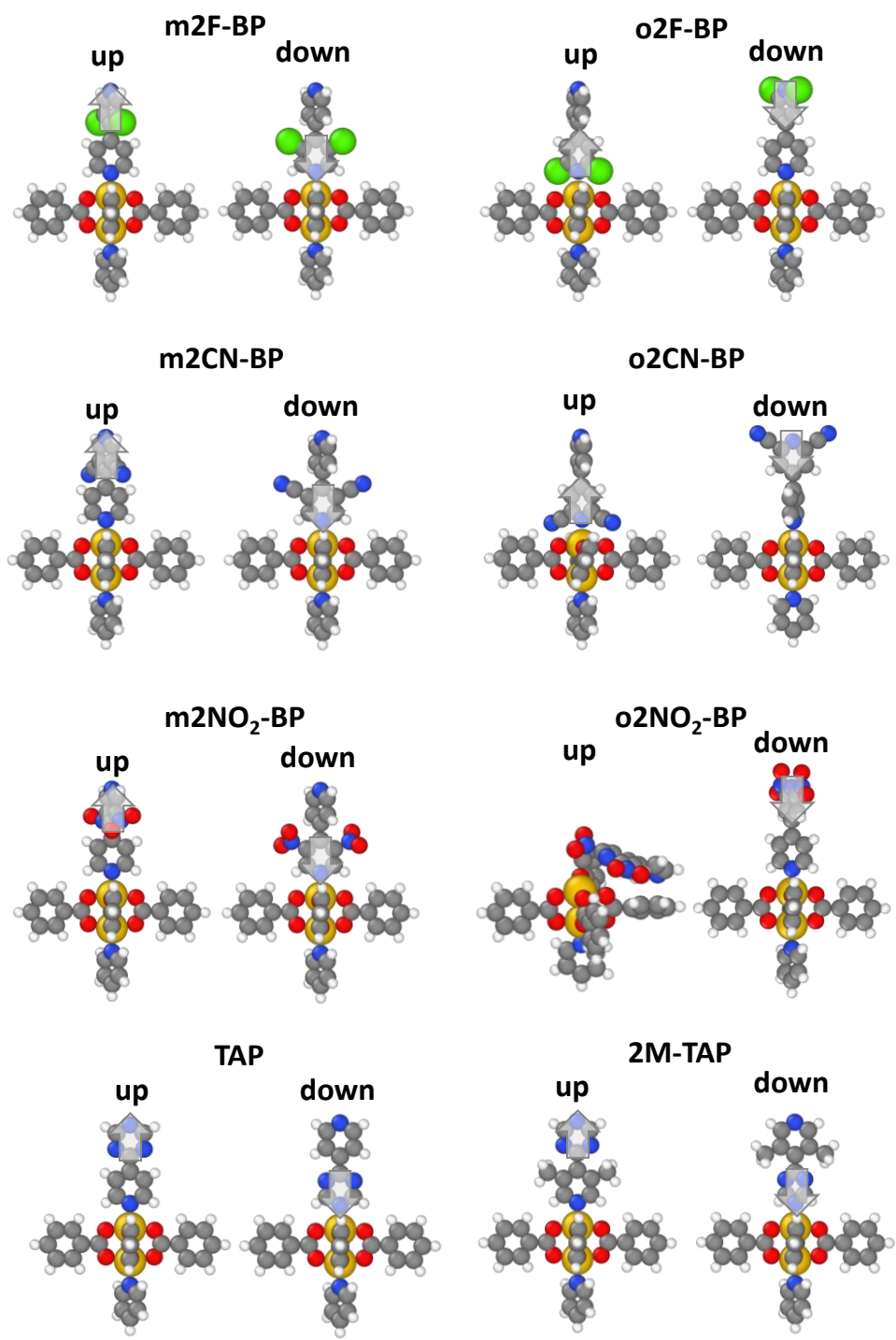

*Figure S6: Structures of the up and down configurations of saturated Zn-paddlewheel monomers bonded to polar apical linkers (yellow: C; cyan: H; dark blue: N; red: O; violet: Zn): m2F-BP: 3,5-difluoro-4,4'-bipyridine; o2F-BP: 2,6-difluoro-4,4'-bipyridine; m2CN-BP: 3,5-dicyano-4,4'-bipyridine; o2CN-BP: 2,6-dicyano-4,4'-bipyridine; m2NO$_2$-BP: 3,5-dinitro-4,4'-bipyridine; o2NO$_2$-BP: 2,6-dinitro-4,4'-bipyridine; TAP: 4-s-triacinylpyridine; 2M-TAP: 3,5-dimetil-4-s- triacinylpyridine linker. C: yellow spheres; O: red spheres; H: small cyan spheres; F: green spheres, Zn: large, purple spheres*

## 6. Dipole per unit cell per polar layer

Assuming that molecular dipoles and bonding dipoles of the monomeric units from Figure S6 can be simply superimposed, one can calculate a rough estimate of the dipole per unit cell per layer of polar apical linkers.

$$\Delta\mu = \mu_{down} - \mu_{up} - \mu_{mol}$$

This can be rationalized by the following considerations: The dipole of $\mu_{down}$ consists of the dipole of the Zn-N bond for the "down" oriented linker plus the molecular dipole; the dipole of $\mu_{up}$ consists of the dipole of the Zn-N bond for the "up" oriented linker and the rotated molecules (i.e., plus -$\mu_{mol}$). When subtracting $\mu_{up}$ from $\mu_{down}$ that corresponds to rotating the structure of the up configuration, which is exactly what one needs to correctly account for the bond dipoles. The molecular dipole is, however counted twice (as -(-$\mu_{mol}$)=+$\mu_{mol}$ from the rotated monomer "up"-configuration. To account for that, $\mu_{mol}$ has to be subtracted to get $\Delta\mu$.

## 7. Comparison of the electronic properties of the m2F-BP and o2F-BP monolayers infinitely extended in the directions perpendicular to the dipole direction

*Table S2: Comparison between the global band gap, $E_{G,global}$, and the work function change $\Delta\Phi$ of the first layer for the two different difluorobipyridine linkers. For the definition of the band gap and the work function change see main text.*

|  | $E_{G,global}$ (eV) | $\Delta\Phi$ (eV) |
|---|---|---|
| 3,5-difluoro-4,4'-bipyridine (m2F-BP) | 1.60 | 0.27 |
| 2,6-difluoro-4,4'-bipyridine (o2F-BP) | 1.39 | -0.29 |

## 8. Check for an activation energy in the bond-formation process

In order to more clearly assess the role of kinetics in the bond-formation process, we also checked, whether an activation energy would have to be overcome during bond formation. To that aim, the energy of the "up" m2F-BP-monomer (Figure S6, top left structure) was evaluated as a function of the distance between the apical linker and the saturated Zn-paddlewheel, starting from the equilibrium distance. The results shown in Figure S7 indicate that there is no activation energy for the bonding process. Qualitatively the same results were obtained for the "down" system.

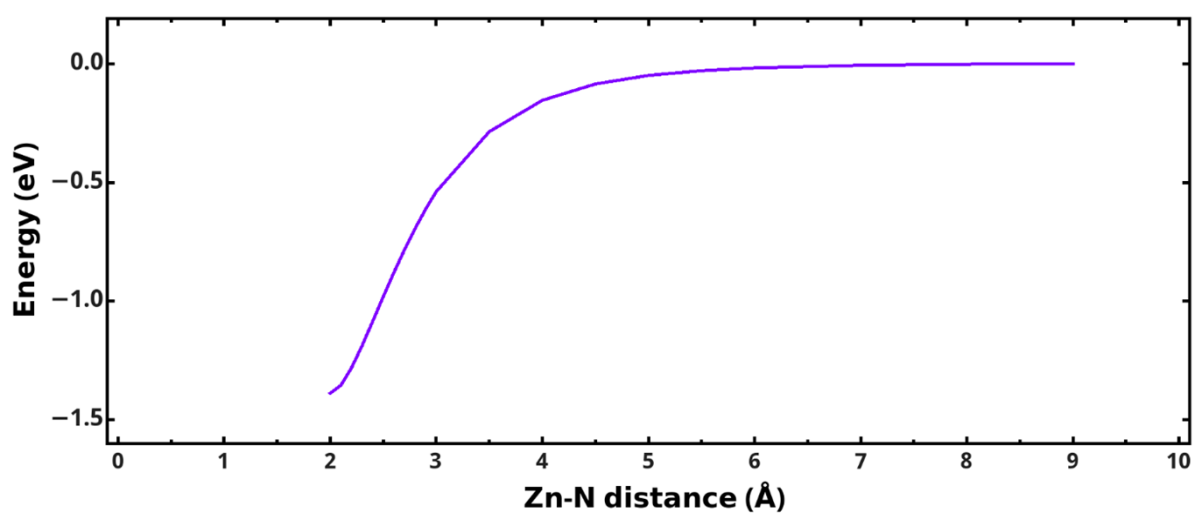

*Figure S7: PBE-calculated total energy of the "up" m2F-BP-monomer as a function of the distance between the Zn-atom of the saturated paddlewheel and the N atom in the apical linker. The curve starts at the equilibrium distance of the bond at 2.08 Å. The energies are given relative to the value at infinite distance and they have been obtained via single-point calculations without optimizing the structures at every distance.*

## 9. Dipole-dipole interactions within monolayers

To check whether the interactions between neighboring dipole units play a role, slab calculations for the three systems depicted in Figure S8 were performed. In the "up" and "down" configurations, the dipole units are parallel. Instead, in the "checkerboard" system the orientation of the dipoles is alternate. Note that the systems studied here consist of only one layer of saturated Zn-paddlewheels connected by BDC linkers to which a monolayer of apical linkers has been bonded. I.e., the top-layer of BDC-linked paddlewheels is missing, as otherwise no insights on the bonding asymmetry could be obtained. The bonding energies are compared in Table S4. Contrary to what one might expect based on dipole-dipole attraction/repulsion arguments, the "checkerboard" structure is not the most stable one. In fact, the bonding energy of that structure is essentially half way between the "up" and "down" configurations. This shows that dipole-dipole interactions play an only negligible role, which can be attributed to the comparably small encountered dipole moments and especially to the large inter-dipole distances of the highly porous systems (inter-dipole distances of about 11Å). This assessment is also backed by simple electrostatic arguments (see main manuscript).

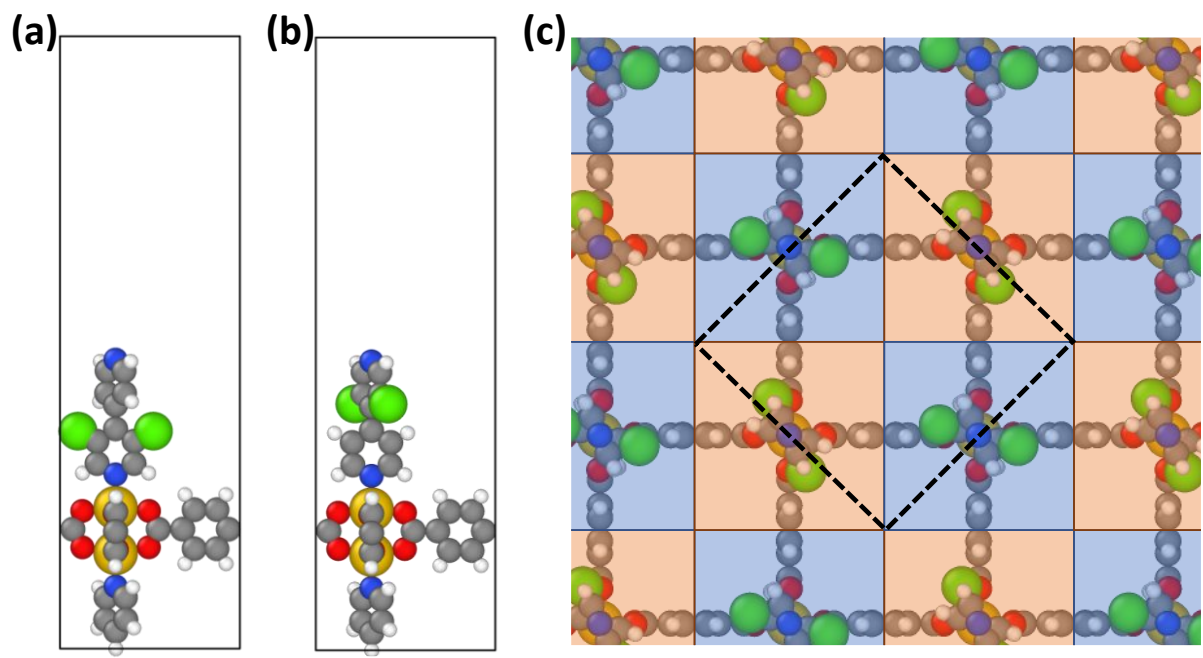

*Figure S8: Systems used for slab calculations, to investigate the interactions between adjacent dipole units. In panels (a) and (b) side-views of the "down" (left) and "up" (right) systems are depicted. Panel (c) shows a top-view of the "checkerboard system", where different shadings indicate dipole-up and down orientations and the dashed black square shows the unit cell (containing on m2F-BP and one o2F-BP molecule). C atoms are depicted in grey, H atoms in white, O atoms in red, N in blue, Zn in yellow, and F in green.*

Table S4: Comparison of the bonding energies, $E_b$, for the "up", "down" and "checkerboard" slab systems infinitely extended in c- and y-directions.

|  | $E_b$ (meV) | $\Delta E_b$ (meV) |
|---|---|---|
| up | 1108 | |
| down | 1048 | -60 meV |
| checkerboard | 1082 | |

## 10. Interaction between polar molecules in neighboring layers

To quantify the interaction between polar molecules in neighboring layers, we performed simulations on dimer clusters as well as on slabs containing two periodic layers of apical linkers. In contrast to the n=2 situation in Figure S1b, only two Zn-paddlewheels (or BDC-linked layers of Zn-paddlewheels) are considered in analogy to the situation for the monolayers shown in Figure S8. This is necessary to be able to calculate bonding asymmetries. Bonding energies are then calculated here for the removal of the topmost m2F-BP molecule or monolayer (where for the latter the energy is given again per apical linker molecule).

An analysis of the bonding energies Table S5 reveals that again, there is virtually no interaction between m2F-BP in neighboring layers, as the differences in bonding energies for open as well as for periodic boundary conditions are essentially identical to the monomer value reported in Table 1 of the main manuscript (-66 meV). This highlights that also here the asymmetry in the difference in bonding energies primarily originates from the orientation of the topmost m2F-BP relative to the Zn-paddlewheel it is bonded to.

Table S5: Comparison of the bonding energies, $E_b$, of the four systems depicted in Figure S9. The reported values describe the energy cost of removing of the topmost m2F-BP molecule/monolayer.

| system | dipole orientation | structure in Figure S9 | $E_b$ (meV) | $\Delta E_b$ (meV) |
|---|---|---|---|---|
| cluster | identical | (a) | 1022 | |
| | opposite | (b) | 1090 | -68 |
| periodic slab | identical | (c) | 1082 | |
| | opposite | (d) | 1141 | -59 |

[1]     V. Blum, R. Gehrke, F. Hanke, P. Havu, V. Havu, X. Ren, K. Reuter, M. Scheffler, *Computer Physics Communications* **2009**, *180*, 2175.

[2]     A. Tkatchenko, M. Scheffler, *Physical Review Letters* **2009**, *102*, DOI 10.1103/PhysRevLett.102.073005.